\documentclass[a4paper,aps,twocolumn,nobibnotes,pra,showpacs]{revtex4-1}

  \usepackage{color}
  \usepackage{graphicx}
  \usepackage{amssymb}
  \usepackage{amsmath}
  \usepackage{bm}
  \usepackage{verbatim}
  \usepackage[utf8]{inputenc}
  \usepackage{bbm}
  	
\newcommand{\ket}[1]{\mbox{$ | #1 \rangle $}}
\newcommand{\bra}[1]{\mbox{$ \langle #1 | $}}
\newcommand{\be}{\begin{eqnarray}}
\newcommand{\ee}{\end{eqnarray}}

\newcommand{\PP}{\ensuremath{\mathcal{P}}}
\newcommand{\QQ}{\ensuremath{\mathcal{Q}}}

\begin{document}

\title{Steering criteria from general entropic uncertainty relations}
\author{Ana C. S. Costa}
\author{Roope Uola}
\author{Otfried G\"uhne}
\affiliation{Naturwissenschaftlich-Technische Fakult\"at, Universit\"at Siegen, Walter-Flex-Stra{\ss}e 3, 57068 Siegen, Germany}
\date{\today}

\begin{abstract}
The effect of steering describes a possible action at a distance via 
measurements but characterizing the quantum states that can be used 
for this task remains difficult. We provide a method to derive sufficient 
criteria for steering from entropic uncertainty relations using 
generalized entropies.  We demonstrate that the resulting criteria 
outperform existing criteria in several scenarios; moreover, they 
allow one to detect weakly steerable states.
\end{abstract}


\maketitle

{\it Introduction.---} 
Steering is a term coined by Schr\"odinger in 1935 in order to capture the
essence of the Einstein-Podolsky-Rosen argument~\cite{schroedingerletter}. 
It describes Alice's ability to affect Bob's quantum state through her choice 
of a measurement basis, without allowing for instantaneous signaling. In the 
modern view, steering is based on a quantum correlation between entanglement
and the violation of Bell inequalities, meaning that not every entangled state 
can be used for steering and not every steerable state violates a Bell 
inequality~\cite{wiseman}.

In the recent years the theory of steering has evolved quickly. It has been 
shown that the concept of steering is closely related to fundamental problems
and open questions in quantum physics. For instance, steering has been used 
to find counterexamples to so-called Peres conjecture, which was an open 
problem in entanglement theory for more than 15 years~\cite{moroder, Vertesi2014, Yu2017}. In addition, steering was shown to be equivalent to the
notion of joint measurability of generalized measurements \cite{Quintino2014,
Uola2014, Uola2015, Uola2017,Kiukas2017} and results from one problem can be transferred
to the other. Finally, steering has been shown to be useful for tasks in 
quantum information processing, such as one-sided device-independent quantum 
key distribution \cite{Branciard2012} and subchannel discrimination \cite{Piani2015}.

Despite all of these results, the simple question whether or not a given bipartite
quantum state is useful for steering is not easy to answer. If
the conditional states of Bob are known, the problem can be solved via semidefinite programming~\cite{pusey, skrypzikreview,Kogias2015}, but this approach requires knowledge of Alice's measurements and is restricted to small dimensions. Other
steering criteria exist~\cite{wiseman,ECavalcanti2015,Jevtic2014,chau2016,Bowles2016,Moroder2016}, 
but general concepts for the derivation
of them are missing. This is in contrast to entanglement theory, where concepts
such as the theory of positive, but not completely positive maps provide a 
guiding line for developing separability criteria~\cite{hororeview}.

In this paper we identify entropic uncertainty relations as a fundamental 
tool to develop steering criteria. Uncertainty relations in terms of 
entropies have already become important in many areas of quantum information 
theory~\cite{Maassen1988, wehnerreview}. We show that various entropic uncertainty
relations can be transformed into steering criteria. As examples, we consider 
generalized entropies such as the so-called Tsallis entropy and demonstrate
that the resulting criteria outperform known steering inequalities in many cases. Our approach
is motivated by previous works on entanglement criteria from entropic uncertainty
relations~\cite{Guhne2004,Huang2010} and
it generalizes recent entropic criteria for steering~\cite{walborn11, schneeloch13}, which were, however, restricted to the special case of the Shannon entropy.


{\it Steering and entropies.---} 
In steering scenarios, one assumes that Alice and Bob share a quantum 
state $\varrho_{AB}$. Then, Alice makes measurements on her system 
and claims that with these measurements she can steer the state 
inside Bob's laboratory. Bob, of course, is not convinced of Alice's 
abilities. 
In a more formal manner, we can assume that Alice performs a measurement 
$A$ with outcome $i$ on her part of the system, while Bob performs a 
measurement $B$ with outcome $j$ on his part. From that, they can obtain
the joint probability distribution of the outcomes. If for all possible
measurements $A$ and $B$ one can express the joint probabilities in the 
form
\be
\label{lhs-model}
p(i,j|A,B) = \sum_\lambda p(\lambda)p(i|A,\lambda)p_q(j|B,\lambda),
\ee
then the system is called unsteerable. Here, $p(i|A,\lambda)$ is a general
probability distribution, while $p_q(j|B,\lambda) = \textrm{Tr}_B[B(j) \sigma_\lambda]$
is a probability distribution originating from a quantum state $\sigma_\lambda$. 
Furthermore, $B(j)$ denotes a measurement operator such that 
$\sum_j B(j) = \mathbbm{1}$, and $\sum_\lambda p(\lambda) = 1$, where $\lambda$ 
is a label for the hidden quantum state $\sigma_\lambda$. A model as in
Eq.~(\ref{lhs-model}) is called a local hidden state (LHS) model, and if it
exists, Bob can explain all the results through a set of local states $\{\sigma_\lambda\}$
which is not altered by Alice's measurements. 
But if it is not possible to find states $\sigma_\lambda$ that make this 
probability distribution feasible, Bob concludes that Alice can steer 
the state. 

Let us now explain some basic facts about entropy. For a general 
probability distribution $\PP = (p_1,\dots,p_N)$, the Shannon 
entropy is defined as~\cite{coverthomas} 
\be
S(\PP) = - \sum_i p_i\ln (p_i).
\ee
Entropic uncertainty relations can easily be explained with an example. 
Consider the Pauli measurements $\sigma_x$ and $\sigma_z$ on a single 
qubit. For any quantum state these measurements give rise to a two-valued probability
distribution and to the corresponding entropy $S(\sigma_k).$
The fact that $\sigma_x$ and $\sigma_z$ do not share a common eigenstate can be
expressed as~\cite{Maassen1988}
\begin{equation}
S(\sigma_x) +  S(\sigma_z) \geq \ln(2), 
\end{equation}
where the lower bound does not depend on the state. 

For our approach, we also need the relative entropy, also known as 
Kullback-Leibler divergence~\cite{coverthomas}, between two 
probability distributions $\PP$ and $\QQ$,
\be
D(\PP||\QQ) = \sum_i p_i \ln \Big(\frac{p_i}{q_i}\Big).
\ee
Two properties are essential: First, the relative entropy is additive for 
independent distributions, that is if $\PP_1,\PP_2$ 
are two probability distributions with the joint distribution 
$\PP(x,y) = \PP_1(x)\PP_2(y)$, and the same for $\QQ_1,\QQ_2$, then one has
\be
\label{eq-add}
D(\PP||\QQ)
= D(\PP_1||\QQ_1) 
+ D(\PP_2||\QQ_2).
\ee
Second, the relative entropy is jointly convex. This means that for two
pairs of distributions $\PP_1, \QQ_1$ and $\PP_2, \QQ_2$ one has
\begin{align}
D[\lambda \PP_1 &+  (1-\lambda)\PP_2||\lambda \QQ_1 + (1-\lambda) \QQ_2] 
\nonumber
\\
& \leq \lambda D(\PP_1||\QQ_1)
+ (1-\lambda)D(\PP_2||\QQ_2).
\label{eq-jointconv}
\end{align}

{\it The main idea.---}
The starting point of our method is the relative entropy between two
distributions, namely
\be
F(A,B) = - D(A\otimes B||A \otimes \mathbb{I}).
\ee
Here, $A\otimes B$ denotes the joint probability distribution 
$p(i,j|A,B)$, which we denote by $p_{ij}$ for convenience; $A$ is the marginal distribution
$p(i|A)$, which we denote by $p_{i}$; and $\mathbb{I}$ is a uniform distribution with $q_j = 1/N$  for all outcomes $j \in \{1,\dots ,N\}$. As the relative entropy is 
jointly convex, $F(A,B)$ is concave in the probability distribution 
$A\otimes B$. Hence, we get directly
\begin{equation}
\label{eq-res1}
F(A,B) = 
- \sum_{ij} p_{ij} \ln\Big(\frac{p_{ij}}{p_i/N}\Big) = S(B|A) - \ln(N),
\end{equation}
where $S(B|A) = S(A,B) - S(A)$ is the conditional entropy. 
On the other hand, considering a product distribution $p(i|A,\lambda)p_q(j|B,\lambda)$ with a fixed $\lambda$ and the property from Eq.~(\ref{eq-add}), 
we have
\be
F^\lambda (A,B) & = & - D[p(i|A,\lambda)||p(i|A,\lambda)] 
- D[p_q(j|B,\lambda)||\mathbb{I}] \nonumber \\
&=& S^\lambda(B) - \ln(N).
\label{eq-res2}
\ee
The term $S^\lambda(B)$ in the right-hand side of this equation depends on probability 
distributions taken from the quantum state $\sigma_\lambda$. 
For a given set of measurements $B_k$, 
such distributions typically obey an entropic uncertainty 
relation 
\begin{equation}
\sum_k S^\lambda(B_k) \geq C_B,
\label{eq-eur}
\end{equation}
where $C_B$ is some entropic uncertainty bound for the observables $B_k$. 
Finally, since $S$ is concave, the same bound holds for convex 
combinations of product distributions $p(i|A,\lambda)p_q(j|B,\lambda)$ from Eq.~(\ref{lhs-model}). Connecting this to Eqs.~\eqref{eq-res1} and~\eqref{eq-res2} we have, 
for a set of measurements $A_k\otimes B_k$,
\be
\label{eq-res3}
\sum_k S(B_k|A_k) \geq C_B,
\ee
which means that any nonsteerable quantum system obeys this relation.
In this way entropic uncertainty relations can be used to derive
steering criteria. The intuition behind these criteria is based on the 
interpretation of Shannon conditional entropy. In Eq.~(\ref{eq-res3}), one can see 
that the knowledge Alice has about Bob's outcomes
is bounded. If this inequality is violated, then the system is steerable, meaning 
that Alice can do better predictions than those allowed by an entropic uncertainty relation.

So far, this criterion is the same as the one in
Ref.~\cite{schneeloch13}, but our proof highlights the three central 
ingredients: First, we needed an additivity relation for independent 
distributions in Eq.~(\ref{eq-add}); second, we needed the state-independent entropic uncertainty relation in Eq.~(\ref{eq-eur}); and
finally we needed the joint convexity of the relative entropy in Eq.~(\ref{eq-jointconv}). 
These properties are not at all specific for the Shannon entropy, so our 
strategy works also for generalized entropies.

{\it Steering criteria for generalized entropies.---}
As a possible generalized entropy, we consider the so-called 
Tsallis entropy~\cite{Havrda1967,Tsallis1988} which depends 
on a parameter $q > 1.$ It is given by
\be
\label{tsallis}
S_q (\PP) = - \sum_i p_i^q \ln_q (p_i),
\ee
where the $q$ logarithm is defined as $\ln_q(x) = ({x^{1-q}-1})/({1-q})$. 
Note that in the limit $q \rightarrow 1$ this entropy converges to the 
Shannon entropy. The generalized relative entropy can be defined as~\cite{Tsallis1998,Furuichi2014}
\be
D_q (\PP||\QQ) = 
-\sum_i p_i \ln_q \Big(\frac{q_i}{p_i}\Big).
\ee
This quantity is jointly convex and obeys the following relation for product 
distributions:
\be
D_q (\PP|| \QQ) &=& D_q(\PP_1||\QQ_1) + D_q(\PP_2||\QQ_2) \nonumber \\
&&+ (q-1)D_q(\PP_1||\QQ_1)D_q(\PP_2||\QQ_2).\nonumber
\ee
The additional term is due to non-additivity of the generalized entropy.

Now we can apply the machinery derived above and consider the 
quantity $F(A,B) = -D_q(A\otimes B||A\otimes \mathbbm{I}).$ It 
follows by direct calculation that if the measurements $B_k$ 
obey an entropic uncertainty relation
\be
\sum_k S_q(B_k)\geq C^{(q)}_B
\ee
then one has the steering criterion
\be
\label{tsc}
\sum_k\Big[S_q(B_k|A_k)+(1-q)C(A_k,B_k)\Big]
\geq C_B^{(q)},
\ee
and violation of it implies steerability of the state. 
Here $S_q(B|A) = S_q(A,B) - S_q(A)$ is the conditional 
entropy~\cite{Furuichi2006} and the additional 
term is given by
\begin{equation}
C(A,B) = \sum_ip_i^q [\ln_q(p_i)]^2 
- \sum_{i,j}p_{ij}^q \ln_q (p_i)\ln_q (p_{ij}).
\end{equation}
From Eq.~(\ref{tsc}) it is easy to see that if we consider $q\rightarrow 1$, we arrive 
at Eq.~(\ref{eq-res3}). Note that we can also rewrite Eq.~\eqref{tsc} 
in terms of probabilities as
\begin{equation}
\label{tsc-prob}
\frac{1}{q-1}\Big[\sum_{k}\big(1 - \sum_{ij}\frac{(p_{ij}^{(k)})^q}{(p_{i}^{(k)})^{q-1}}\big)\Big] \geq C_B^{(q)}.
\end{equation}
Here, $p_{ij}^{(k)}$ is the probability of Alice and Bob for outcome $(i,j)$ 
when measuring $A_k\otimes B_k$, and $p_i^{(k)}$ are the marginal outcome probabilities of Alice's measurement $A_k$. This form of the criterion
is straightforward to evaluate.

{\it Application I: Isotropic states.---} 
To test the strength of our steering criteria we consider $d$-dimensional 
isotropic states~\cite{Horodecki1999}
\be
\varrho_{\rm iso} = 
\alpha\ket{\phi^+_d}\bra{\phi^+_d} + \frac{1-\alpha}{d^2}\mathbbm{1},
\ee
where $\ket{\phi^+} = ({1}/{\sqrt{d}})\sum_{i=0}^{d-1} \ket{i}\ket{i} $ is a maximally 
entangled state. 
These states are known to be entangled for $\alpha > 1/(d+1)$ and separable otherwise. As observables, we consider $m$ mutually unbiased bases 
(MUBs) in dimension $d$ (provided that they exist). One can directly check that the marginal probabilities 
for this class of states are $p_i = 1/d$ for all $i$ and the joint probabilities are $p_{ii} = [1+(d-1)\alpha]/d^2$ (occurring $d$ times), and $p_{ij} = (1-\alpha)/d^2$ [for $i \neq j$ and occurring $d(d-1)$ times]. These 
probabilities are independent of the chosen measurements. Inserting them 
in Eq.~(\ref{tsc-prob}), the condition for non steerability reads
\begin{equation}
\label{eq-iso}
\frac{m}{q-1}
\big(1 - \frac{1}{d^q}\{[1+(d-1)\alpha]^q + (d-1)(1-\alpha)^q\}\big)
\geq C_B^{(q)},
\end{equation}
which depends on the parameter $q$ and the number of MUBs $m$. For 
certain values of $q$ and $m$, the bounds of the entropic uncertainty 
relations $C_B^{(q)}$ are known (see the Appendix). For other cases they can be approximated numerically.

Let us discuss the strength of this criterion. First, numerical
investigations suggest that the criterion is strongest for $q=2.$
For this value of $q$ the violation of Eq.~(\ref{eq-iso})
occurs for $\alpha > 1/\sqrt{m}$. Considering a complete set of MUBs 
$(m = d+1)$ (this exists for $d$ being a power of a prime) the violation 
happens for $\alpha > 1/\sqrt{d+1}$.

For qubits ($d=2$) isotropic states are equivalent to Werner states~\cite{Werner1989}. Then, with a complete set of MUBs the violation occurs for 
$\alpha > 1/\sqrt{3} \approx 0.577$, which is known to be the 
optimal threshold~\cite{Cavalcanti2009}. More generally, in Ref.~\cite{Cavalcanti2015}, a steering inequality for MUBs and isotropic states has been presented which is violated
for $\alpha > (d^{3/2}-1)/(d^2 -1)$. It is straightforward to show that our inequality is stronger. Recently, the same problem has been investigated 
using semidefinite programming~\cite{Bavaresco2017}. For 
$3\leq d \leq 5$ a better threshold than ours was obtained, but 
it is worth mentioning that our criteria directly use probability 
distributions from few measurements, without the need of performing 
full tomography on Bob's conditional state. In addition, numerical 
approaches are naturally limited to small dimensions.

\begin{figure}
\includegraphics[scale=0.56]{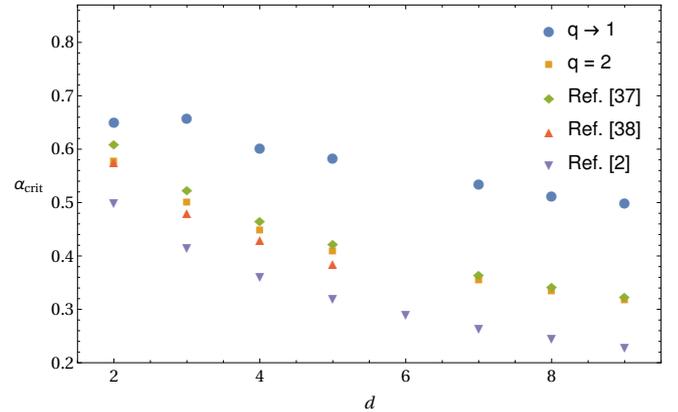}
\caption{(Color online) The critical value of white noise $\alpha$ for 
different dimensions $d$, considering a complete set of MUBs. In this 
plot, blue circles correspond to our criterion in Eq.~(\ref{eq-iso})
for $q\rightarrow 1$ and the yellow squares to $q=2$. The green diamonds 
correspond to the results for the inequality presented in Ref.~\cite{Cavalcanti2015} and the red triangles in Ref.~\cite{Bavaresco2017}, where $\alpha_{\textrm{crit}}$ was calculated via semidefinite programming (numerical method). Below the purple reversed triangles the existence of an LHS model for all projective measurements (i.e. infinite amount of measurements instead of $d+1$ MUBs) is known~\cite{wiseman}. Note that Ref. [2] is given for comparison; this is not a steering criterion, but a bound on any criterion.}
\label{fig1}
\end{figure}

In Fig.~\ref{fig1}, we compare our criterion with the ones mentioned 
above. We concentrate on the values of $q \rightarrow 1$ and $q=2$, since the former is related to the usual entropic steering criteria and the latter is the optimal value of $q$ for the detection of steerable states.

{\it Connection to existing entanglement criteria.---}
At this point, it is interesting to compare our approach with
entanglement criteria derived from entropic uncertainty
relations~\cite{Guhne2004}. The mathematical formulation goes as follows. Let 
$A_1$ and $A_2$ ($B_1$ and $B_2$) be observables on Alice's (Bob's) 
laboratory. Assume that Bob's observables obey an entropic uncertainty
relation $S(B_1) + S(B_2) \geq C_B$, where $S(B_i)$ is a generalized 
entropy, such as the Shannon or Tsallis entropy. Then it can be shown 
that for separable states
\be
\label{eq-criteria-sep}
S(A_1\otimes B_1) + S(A_2\otimes B_2) \geq C_B
\ee
holds. Here, $S(A_k\otimes B_k)$ is the entropy of the probability
distribution of the outcomes of the {\it global} observable 
$A_k\otimes B_k$. Note that for a degenerate
$A_k\otimes B_k$ the probability distribution differs from the 
local ones. For instance, measuring $\sigma_z \otimes \sigma_z$
gives four possible local probabilities $p_{++}, p_{+-}, p_{-+}, p_{--},$
but for the evaluation of $S(A_k\otimes B_k)$ one combines them
as $q_+ = p_{++}+p_{--}$ and  $q_- = p_{+-}+p_{-+}$, as these
correspond to the global outcomes. 

Some connections to our derivation of steering inequalities are interesting. 
First, if one reconsiders
the proof in Ref.~\cite{Guhne2004} one realizes that Eq.~(\ref{eq-criteria-sep})
can actually be extended to a steering criterion. That is, 
all probability distributions of the form in Eq.~(\ref{lhs-model}) fulfill it.
Second, also in Ref.~\cite{Guhne2004} it was observed that the criterion is
strongest for values $2 \leq q \leq 3.$ Third, if one asks for a direct comparison between Eq.~(\ref{eq-criteria-sep}) and Eqs.~(\ref{tsc}) and (\ref{eq-res3}) 
one finds that Eq.~(\ref{eq-criteria-sep}) is of the same 
strength for special scenarios (e.g. Bell-diagonal two-qubit states and Pauli
measurements), while it seems weaker in the general case (see below). Finally, note
that the approach of Ref.~\cite{Guhne2004} has been slightly improved in Ref.~\cite{Huang2010}, and the resulting criteria can also be extended into steering 
inequalities.


{\it Application II: General two-qubit states.---}
Let us now consider the application of our methods to general two-qubit 
states. Any two-qubit state can, after application of local unitaries, be
written as
\begin{equation}
\label{eq-rho}
\varrho_{AB} =
\frac{1}{4}
\big[\mathbbm{1}\otimes\mathbbm{1}+
(\vec{a}\vec{\sigma})\otimes\mathbbm{1}
+\mathbbm{1}\otimes(\vec{b}\vec{\sigma})
+\sum_{i=1}^3 c_i \sigma_i\otimes \sigma_i\big]
\end{equation}
where $\vec{a},\vec{b},\vec{c} \in\mathbb{R}^3$ are vectors with 
norm not higher than 1, $\vec{\sigma}$ is a vector composed of the Pauli matrices, and $(\vec{a}\vec{\sigma})= \sum_i a_i \sigma_i$. 
Let us assume that Alice performs projective measurements with 
effects $P_k^A = [\mathbbm{1} + \mu_k (\vec{u}_k\vec{\sigma})]/2$ 
and Bob with the effects 
$P_k^B = [\mathbbm{1} + \nu_k (\vec{v}_k\vec{\sigma})]/2$ with $\mu_k, \nu_k = \pm 1$
and $\{\vec{u},\vec{v}\}\in\mathbb{R}^3$. 
Then, Eq.~\eqref{tsc-prob} can be written as
\begin{align}
\sum_k
\Big[
1 
- &
\sum_{\mu_k, \nu_k}
\frac{[1 
+ \mu_k(\vec{a}\vec{u}_k) 
+ \nu_k(\vec{b}\vec{v}_k) 
+ \mu_k\nu_k T_{k}]^q}
{2^{q+1}[1+\mu_k(\vec{a}\hat{u}_k)]^{q-1}}
\Big]
\nonumber
\\
&\geq (q-1) C_B^{(q)}, 
\end{align}
where $T_{k}=\sum_{i=1}^3 c_i u_{ik}v_{ik}$. The optimization 
over measurements of this criterion for general two-qubit states 
is involving. We will focus on the simple case of Pauli measurements,
meaning that $\vec{u}_k = \vec{v}_k = \{(1,0,0)^T, (0,1,0)^T,(0,0,1)^T\}$ 
and $q=2$. Then we have the following inequality:
\begin{equation}
\label{criteria-general}
\sum_{i=1}^3\left[\frac{1 - a_i^2 - b_i^2 - c_i^2 + 2a_ib_ic_i}{2(1-a_i^2)}\right] \geq 1,
\end{equation}
the violation of which implies steerability.

Now, we can compare our criteria with other proposals for the detection 
of steerable states using three measurements. 
The criteria from Eq.~\eqref{eq-criteria-sep} prove steerability 
if $\sum_{i=1}^3 c_i^2 > 1$, and from the linear criteria~\cite{wiseman, Costa2016} steerability follows if $({\sum_{i=1}^3 c_i^2})^{1/2} > 1$, which is equivalent (see Appendix B).  
Not surprisingly, Eq.~(\ref{criteria-general}) is stronger, 
since it uses more information about the state. This statement can be made
hard by analyzing $10^6$ Hilbert-Schmidt random two-qubit states (i.e., partial states from a uniform distribution of pure states on a larger system)~\cite{random-ensemble}. 
$94.34\%$ of the 
states do not violate any of the criteria, $3.81\%$ are steerable according 
to all three criteria mentioned above, $1.85\%$ violate only criterion~\eqref{criteria-general}, 
and no state violates only the linear criteria.

A special case of two-qubit states is the Bell diagonal ones, which 
can be obtained if we set $\vec{a} = \vec{b} = 0$ 
in Eq.~\eqref{eq-rho}. For this class of states it is easy to see that the 
three criteria are equivalent. Note, however, that a necessary 
and sufficient condition for steerability of this class
for projective measurements has recently been found~\cite{chau2016}.

{\it Application III: One-way steerable states.---}
As an example of weakly steerable states that can be detected with 
our methods we consider one-way steerable states, i.e., states that 
are steerable from Alice to Bob and not the other way around. We 
consider the state
\begin{equation}
\label{one-way-state}
\varrho_{AB} = \beta\ket{\psi (\theta)}\bra{\psi (\theta)}+(1-\beta)\frac{\mathbbm{1}}{2}\otimes \varrho_B^\theta ,
\end{equation}
where $\ket{\psi (\theta)} = \cos (\theta)\ket{00} + \sin (\theta)\ket{11}$ 
and $\varrho_B^\theta = \text{Tr}_{A}[\ket{\psi (\theta)}\bra{\psi (\theta)}]$. 
It has been shown  that for $\theta \in [0,\pi/4]$ and 
$\cos^2 (2\theta) \geq ({2\beta-1})[{(2-\beta)\beta^3}]$ this state 
is not steerable from Bob to Alice considering all possible projective measurements~\cite{Bowles2016}, while Alice can steer 
Bob for $\beta > 1/2$. 

Considering three measurement settings, this 
state is one-way steerable for 
${1}/{\sqrt{3}} < \beta \leq \beta_{\rm max}$
with $\beta_{\rm max}= [{1+2\sin^2(2\theta)}]^{-1/2}$ \cite{Xiao2017}. 
For our entropic steering criteria we consider three Pauli 
measurements and $q=2$ and we find that this state is one-way 
steerable for
\be
\label{esc-one-way}
\frac{1}{2\cos(2\theta)}\sqrt{3-\sqrt{1+8\sin^2 (2\theta)}}
< 
\beta 
\leq \beta_{\rm max}. 
\ee
For any $\theta$ this gives a non empty interval of $\beta$ for which our criterion detects these
weakly steerable states. An attempt at optimizing over the set of measurements 
will be addressed in a future work.


{\it Conclusions.---} 
In this work we have proposed a straightforward technique for the 
construction of strong steering criteria from entropic uncertainty relations. 
These criteria are easy to implement using a finite  set of measurement 
settings only, and do not need the use of semidefinite programming and 
full tomography on Bob's conditional states. 

For future work, several directions seem promising. First, besides the 
usual entropic uncertainty relations, such as entropic uncertainty 
relations in the presence of quantum memory~\cite{Berta2010} or relative
entropy formulations of the uncertainty principle~\cite{Toigo2017} are promising 
starting points for other criteria. Second, one can try to make quantitative
statements on steerability from steering criteria. Recently, some attempts
in this direction have been pursued~\cite{Schnee2017}. Also recently, a proposal
for multipartite steering criteria based on Shannon entropy has been proposed by Riccardi {\it et al.}~\cite{Riccardi2017}. Finally, it would
be highly desirable to embed our approach in a general theory of multiparticle
steering~\cite{Riccardi2017}. 

{\it Acknowledgments.---}
We thank Marcus Huber and Renato M. Angelo for discussions. This work was 
supported by the DFG, the ERC (Consolidator Grant No. 683107/TempoQ) 
and the Finnish Cultural Foundation.


\section*{Appendix}
\subsection*{A: Known entropic uncertainty relations}

Here we present different entropic uncertainty relations 
that were used in this work and known from literature. For the Shannon entropy ($q\rightarrow 1$) 
and a complete set of MUBs, entropic uncertainty relations were 
analytically derived in~Ref.~\cite{Ruiz1995} and are given by
\be
\label{boundq1}
C_B = 
\begin{cases}
(d+1)\ln \left(\frac{d+1}{2}\right), & \quad d\, \text{odd} \\ \\
\frac{d}{2}\ln \left(\frac{d}{2}\right) + \left(\frac{d}{2}+1\right)\ln \left(\frac{d}{2}+1\right), & \quad d\, \text{even}. \\
\end{cases}
\ee
For the Tsallis entropy and $m$ MUBs it has been shown in Ref.~\cite{Rastegin2013} that, for $q\in (0;2]$, the bounds 
are given by 
\be
\label{boundq2}
C_B^{(q)} = m\ln_q \left(\frac{m d}{d + m -1}\right).
\ee
If we consider the case $q\rightarrow 1$, this bound is not optimal 
for even dimensions, so in this case it is more appropriate to consider 
the bounds given in Eq.~\eqref{boundq1}.

\subsection*{B: Details on two-qubit calculations}

First, consider the steering  criterion in Eq.~(\ref{eq-criteria-sep}),
developed in Ref.~\cite{Guhne2004}. For three Pauli measurements and 
the Tsallis entropy, we have the following relation
\be
\sum_{k=1}^3 S_q (A_k\otimes B_k) \geq C_B^{(q)},
\ee
where $A_k = (\vec{u}_k\vec{\sigma})$ and 
$B_k = (\vec{v}_k\vec{\sigma})$. In terms of probabilities 
this criterion can be rewritten as
\begin{align}
&\frac{1}{q-1}\sum_{k=1}^3
\Big\{1-\Big[p_{\vec{u}_k,\vec{v}_k}(+1,+1) + p_{\vec{u}_k,\vec{v}_k}(-1,-1)\Big]^q \nonumber \\
&- \Big[p_{\vec{u}_k,\vec{v}_k}(+1,-1) + p_{\vec{u}_k,\vec{v}_k}(-1,+1)\Big]^q\Big\} \geq C_B^{(q)}.
\end{align}
Inserting the probabilities for general two-qubit systems, we have that
\begin{equation}
\frac{1}{q-1}\sum_{k=1}^3\Big\{1- 2^{-q}\Big[(1+T_{k})^q + (1-T_{k})^q\Big]\Big\} \geq C_B^{(q)}.
\end{equation}
If we fix the measurements and the value of $q$ in the same way as in Eq.~\eqref{criteria-general}, this criterion gives 
$\sum_{i=1}^3 c_i^2 \leq 1$. Then, if this inequality is violated,
the system is steerable.


\end{document}